\documentclass{sig-alternate-2013}
\usepackage{graphicx}
\usepackage{amsmath}
\usepackage{multirow}

\usepackage{url}
\usepackage{float}
\usepackage{color}
\usepackage{enumerate}

\newcommand{\ainote}[1]{\textcolor{blue}{\small [AI: #1]}}

\newcommand{\ignore}[1]{}

\newcommand{\ya}{\textsc{\emph{YA}}}

\newfont{\mycrnotice}{ptmr8t at 7pt}
\newfont{\myconfname}{ptmri8t at 7pt}
\let\crnotice\mycrnotice%
\let\confname\myconfname%

\permission{Permission to make digital or hard copies of all or part of this work for personal or classroom use is granted without fee provided that copies are not made or distributed for profit or commercial advantage and that copies bear this notice and the full citation on the first page. Copyrights for components of this work owned by others than ACM must be honored. Abstracting with credit is permitted. To copy otherwise, or republish, to post on servers or to redistribute to lists, requires prior specific permission and/or a fee. Request permissions from Permissions@acm.org.}
\conferenceinfo{HT'15,}{September 1--4, 2015, Guzelyurt, TRNC, Cyprus.} 
\copyrightetc{\copyright~2015 ACM. ISBN \the\acmcopyr}
\crdata{978-1-4503-3395-5/15/09\ ...\$15.00.\\
http://dx.doi.org/10.1145/2700171.2791034}

  \makeatletter
  \let\@copyrightspace\relax
  \makeatother
\begin{document}
%
% --- Author Metadata here ---
%\conferenceinfo{WebSci}{'14 Bloomington, Indiana USA}`
%\CopyrightYear{2007} % Allows default copyright year (20XX) to be over-ridden - IF NEED BE.
%\crdata{0-12345-67-8/90/01}  % Allows default copyright data (0-89791-88-6/97/05) to be over-ridden - IF NEED BE.
% --- End of Author Metadata ---

%\title{The Social Ecosystem of Yahoo! Answers: Friends, Activities and Abuses}
%\title{Deviance and Retention in Yahoo! Answers}

\title{Cultures in Community Question Answering}
\subtitle{[Please cite the Hypertext'15 version of this paper]}
%Format\titlenote{(Produces the permission block, and
%copyright information). For use with
%SIG-ALTERNATE.CLS. Supported by ACM.}}
%\subtitle{[Extended Abstract]
%\titlenote{A full version of this paper is available as
%\textit{Author's Guide to Preparing ACM SIG Proceedings Using
%\LaTeX$2_\epsilon$\ and BibTeX} at

\numberofauthors{5} %  in this sample file, there are a *total*
\author{
% 1st. author
\alignauthor
Imrul Kayes\\
       \affaddr{University of South Florida}\\
       \affaddr{Tampa FL, USA}\\
       \email{imrul@mail.usf.edu}
\and
Nicolas Kourtellis\\
       \affaddr{Telefonica Research}\\
       \affaddr{Barcelona, Spain}\\
       \email{nicolas.kourtellis@telefonica.com}
\and
Daniele Quercia\\
       \affaddr{University of Cambridge}\\
       \affaddr{Cambridge, United Kingdom}\\
       \email{dquercia@acm.org}
\and
Adriana Iamnitchi\\
   	\affaddr{University of South Florida}\\
    	\affaddr{Tampa FL, USA}\\
       	\email{anda@cse.usf.edu}
\and
Francesco Bonchi\\
       \affaddr{Yahoo Labs}\\
       \affaddr{Barcelona, Spain}\\
       \email{bonchi@yahoo-inc.com}
}

\maketitle

\begin{abstract}

%In the real world, \ainote{as opposed to what, dream world?} 
%Social dynamics such as verbal communication are influenced by social norms and values.
%These norms and values vary among countries and  common wisdom tells that online Community Question Answering (CQA) services should not be an exception, as offline behavior is often reflected in online activities.

CQA services are collaborative platforms where users ask and answer questions.
We investigate the influence of national culture on people's online questioning and answering behavior. 
For this, we analyzed  a sample of 200 thousand users in Yahoo Answers from 67 countries.
We measure empirically a set of cultural metrics defined in Geert Hofstede's \emph{cultural dimensions} and Robert Levine's \emph{Pace of Life}  and show that behavioral cultural differences exist in community question answering platforms.
We find that national cultures differ in Yahoo Answers along a number of dimensions such as temporal predictability of activities, contribution-related behavioral patterns, privacy concerns, and  power inequality. 
%(e.g., asking and answering), answering and

%A category including the fourth, optional field follows...
 %[Computers and Society]: Social IssuesÑAbuse and crime
%involving computers;

%\terms{Measurement, Human Factors}


\end{abstract}

\section{Introduction}\label{sec:intro}

%Culture is a collective behavior that shapes how people  communicate their thoughts, feelings, and values~\cite{william2002handbook}.
Cultural differences exist in almost all aspects of social interactions. %verbal and non-verbal communication.
For example, in some cultures in Asia it may be considered disrespectful for people to express their opinions or ask  questions to authority figures (e.g., teachers, elders).
In other cultures (such as USA or Canada) asking questions is expected or even encouraged.

Cross-country cultural variations have been studied %by Geert Hofstede and Robert V. Levine 
in the real world via small-scale experiments and opinion surveys.
Geert Hofstede~\cite{hoftede2010cultures} administered opinion surveys to a large number of IBM employees from different countries in the 1960s and 1970s. 
He discovered five cultural dimensions (individualism, power distance, uncertainty avoidance, masculinity, and long term orientation), that can be attributed to the existence of cultural variations. 
Three of these dimensions, individualism, power distance, and uncertainty avoidance, have been used to assess cultural differences in online contexts such as Twitter communication~\cite{garcia2014twitter}, emoticon usage~\cite{park2014cross} and online scheduling~\cite{reinecke2013doodle}.
In brief, individualism reflects the extent to which an individual is integrated into a group (e.g., individualistic cultures like USA emphasizes mostly on their individual goals, as opposed to collectivist cultures like China that emphasizes on group harmony and loyalty).
Power distance is the extent to which the less powerful members of an organization or society expect and accept that power is distributed unequally (e.g., in high power distance countries subordinates simply comply with their leaders). 
Uncertainty avoidance defines the extent to which society members feel uncomfortable with uncertainty and ambiguity (e.g., the stereotypical Swiss plans everything ahead supposedly to avoid uncertainty).

Psychologist Robert V. Levine~\cite{levine1999pace} proposed the \emph{Pace of Life} metric based on the walking speed of city people over a distance of 60 feet, the service time for standard requests for stamps, and the clock accuracy of city banks.
During the 1990s, Levine employed 19 experimenters in large cities from 31 countries and computed country-specific Pace of Life ranks. 
He found significant differences in Pace of Life across cultures and ranked the cultures based on that. 

Such cross-cultural variations that sociologists and psychologists already found in the offline world lead to our main research question: Does national culture determine how we participate in online Community Question Answering (CQA) platforms?
CQAs such as Yahoo Answers (\ya), Quora and Stack Overflow have been  popular in the last decade.
These platforms are rich and mature repositories of user-contributed questions and answers.
For example, \ya, launched in December 2005, has more than one billion posted answers,\footnote{http://www.yanswersbloguk.com/b4/2010/05/04/1-billion-answers-served/} and Quora, one of the fastest growing CQA sites has seen three times growth in 2013.\footnote{http://www.goo.gl/MfK83y}

National cross-cultural variations have been studied in a number of online contexts, including social networks (e.g., Twitter~\cite{golder2011diurnal}, Facebook~\cite{Quercia2013DWH}), location search and discovery (e.g., Foursquare~\cite{silva2014you}) and online scheduling (e.g., Doodle~\cite{reinecke2013doodle}). 
While CQA platforms have been intensively studied ~\cite{Shah2010Answers,Dearman2010Why,Qu2009PQR,Kayes2015WWW}, to the best of our knowledge, there has been no study focusing on users' cultural differences and how they shape asking, answering, or reporting abuses in CQA platforms.
If cultural variations exist in CQA platforms, they could be used for more informed system design, including question recommendation, follow recommendation, and targeted ads.

To fill this gap, we analyzed about 200 thousand sampled \ya\ users from 67 countries who were active between 2012 and 2013.
We tested a number of  hypotheses associated with Hofstede's cultural dimensions and Levine's Pace of Life.
Our results show that \ya\ is not a homogeneous subcultural community: considerable behavioral differences exist between the users from different countries.
We find that users from individualistic countries provide more answers, have higher contribution than take away, and are more concerned about their privacy than those from collective cultures.
Users from individualistic countries are also less likely to provide an answer that violates community norms.
We also find that higher power distance countries  show more indegree imbalance in following relationships compared to lower power distance countries.
Finally, we find that users from higher Pace of Life and lower uncertainty avoidance countries have more temporally predictable activities. %, i.e., asking and answering questions.

The rest of the paper is structured as follows. 
Section~\ref{sec:related} discusses previous analysis of CQA platforms and the existing body of work on online cultural variations.
Section~\ref{sec:ya-details-datasets} presents the \emph{YA} functionalities relevant to this study and the dataset used.
We introduce the hypotheses and present the  results relating to Levine's Pace of Life and Hofstede's cultural dimensions  in \ya\ in Section~\ref{sec:pace-of-life} and Section~\ref{sec:hofstede-culture}, respectively.
We discuss the impact of these results in Section~\ref{sec:discussion}.

%\vspace{-4mm}
\section{Related Work}\label{sec:related}

%A number of study has analyzed cross-country difference of online behavior.
%In this section,  we provide an overview of cross-country difference of online behavior and existing research in CQA platforms. 

%\textbf{Cross-culture mode, emoticon and communication on Twitter.} 
Golder and Macy~\cite{golder2011diurnal} studied collective mood in Twitter across countries from 509 million Twitter posts by 2.4 million users over a 2-year period.
Despite having different cultures, geographies, and religions, all countries (USA, Canada, UK, Australia, India, and English-speaking Africa) in their study showed  similar mood rhythms---people tended to be more positive on weekends and early in the morning.
Park et al.~\cite{park2014cross} examined the variation of Twitter users' emoticon usage patterns in cross cultures.
They used Hofstede's national culture scores of 78 countries and found that  collectivist cultures favor vertical and eye-oriented emoticons, where people within individualistic cultures favor horizontal and mouth-oriented emoticons.
Hofstede's cultural dimensions have also been used to study whether culture of a country is associated with the way people use Twitter~\cite{garcia2013cultural}.
In another study on cross-country Twitter communication, Garcia et al. showed that  cultural variables such as Hofstede's indices, language and intolerance have an impact on Twitter communication volume~\cite{garcia2014twitter}. %\\

%\noindent

%\textbf{Cultural boundaries on Foursquare.} 
Silva et al.~\cite{silva2014you} used  food and drink check-ins in Foursquare to identify cultural boundaries and similarities across populations.
They showed that online footprints of foods and drinks are good indicators of cultural similarities between users, e.g., lunch time is the perfect time for Brazilians  to go for slow food places more often, whereas Americans and English people go for slow foods more at dinner time. 
Extracted features like these allow them to apply simple clustering algorithms such as K-means to draw cultural boundaries across the countries. %\\

%\noindent

%\textbf{Cross-culture happiness in Facebook.} 
Quercia~\cite{Quercia2013DWH} used \emph{Satisfaction With Life} tests and measured happiness of 32,787 Facebook users from 12 countries (Australia, Canada, France, Germany, Ireland, Italy, New Zealand, Norway, Singapore, Sweden, UK, USA ).
He found that despite comparative economic status, country-level happiness significantly varies across the  countries and that it strongly correlates with official well-being scores. %\\

%\noindent

%\textbf{Cultures in online scheduling.} 
Reinecke et al.~\cite{reinecke2013doodle} used about 1.5 million Doodle  polls from 211 countries and territories and studied the influence of national culture on people's scheduling behavior.
Using Hofstede's cultural dimensions, they found that Doodle poll participants from collectivist countries find more consensus than those from predominantly individualist societies. %\\

%\noindent

%\textbf{CQA research.} 
CQA platforms have also attracted much research interest focusing on content, user behavior and applications.
%from diverse communities as information science, HCI and information retrieval. 
%The research mostly focuses on content, user behavior and applications.
Research on CQA  content has analyzed textual aspects  of questions and answers. 
Researchers have proposed algorithmic solutions to automatically determine the quality of questions~\cite{Li2012Question} and answers~\cite{Shah2010Answers}.
%, the extent to which certain questions are easy to answer~\cite{dror2013will}, and the type of a given question (e.g., factual or conversational)~\cite{harper2009facts}. 
Research on CQA user behavior has  been mostly about understanding why users contribute content: that is, why users ask questions (askers are failed searchers, in that, they use CQA sites when web search fails~\cite{Liu2012WSF}); and why they don't answer questions (e.g., they refrain from answering sensitive questions to avoid being  reported for abuse and potentially lose access to the community~\cite{Dearman2010Why}). 
As for applications, researchers have proposed effective ways of recommending questions to the most appropriate answerers ~\cite{Qu2009PQR}, automatically answering questions based on past answers~\cite{Shtok2012LPA}, and retrieving factual answers~\cite{Bian2008Finding} or factual bits within an answer~\cite{Weber2012ALE}. 
Our previous work~\cite{Kayes2015WWW} used  user-provided rule violation reports and user social network features to detect the content abusers in \ya.%\\

However, there has been no  empirical cross-cultural analysis of CQA platforms.
This paper is a first step in this direction and it verifies whether cultural differences are manifested in one such platform,  \ya.

\section{Yahoo Answers}\label{sec:ya-details-datasets}
%======================================

After $9$ years of activity, \ya\ has $56$M monthly visitors (U.S. only).\footnote{http://www.listofsearchengines.org/qa-search-engines}
The functionalities of the \ya\ platform and the dataset used in this analysis are presented next.

%-----------------------------------------------
\subsection{The Platform}
%-----------------------------------------------
%\ainote{things to define/explain:\\
%-- user contributed content: answers, questions, comments;\\
%-- points and the incentive mechanism; user levels and what comes with them;\\
%-- self policing: flags. Who can apply them; what restrictions; how are they validated; user levels vs. validation; editors; consequences\\
%-- social network and update dissemination;\\
%-- }

\emph{YA} is a CQA platform in which community members ask and answer questions on various topics.
Users ask questions and assign them to categories selected from a predefined taxonomy, e.g., \emph{Business \& Finance}, \emph{Health}, and  \emph{Politics \& Government}.
YA has about $1300$ categories.
Users can find questions by searching or browsing through this hierarchy of categories.
A question has a title (typically, a short summary of the question), and a body with additional details. %), see an example question and an answer in Figure~\ref{fig:sample_question_answer}.

\ignore{
\begin{figure}[htbp]
\centering
\includegraphics[height=5cm]{sample_question_answer.pdf}
\caption{An answer (truncated and selected as best)  for a question.}
\label{fig:sample_question_answer}
\end{figure}
}

A user can answer any question but can post only one answer per question.
Questions remain open for four days for others to answer.
However, the asker can select a best answer before the end of this 4-day period, which automatically \emph{resolves} the question and archives it as a \emph{reference} question.
The best answer can also be rated between one to five, known as \emph{answer rating}.
If the asker does not choose a best answer, the community selects  one through voting.
% \ainote{how? assign points within a range? or what? IK: Yes the community can vote. The answer, which gets maximum community votes, is selected as the best answer.}
The asker can extend the answering duration for an extra four days.
The questions left unanswered after the allowed duration are deleted from the site.
%\ainote{really? aren't there questions unanswered after 8 days? IK: Yes true. (A relevant comment: one of the problems QA communities are facing is that even for that extended period, 25 to 60 percent questions remain unanswered)}
In addition to questions and answers, users can contribute comments to questions already answered and archived.
%\ainote{need more details. I don't understand the following: These comments are added by users on answers when the corresponding question of the answers is added as a reference question (the case when best answer is found). IK: When an answer is selected as a best answer, the question and answers remain in the community as an archive. Later, users can�t add answers to that questions but add comments. However, the number of these comments are very low and only 0.07\% of abuse reports on those comments.}

\ya\ has a system of points and levels to encourage and reward participation.\footnote{https://answers.yahoo.com/info/scoring\_system}
A user is penalized five points for posting a question, but if she chooses a best answer for her question, three points are given back.
A user who posts an answer receives two points; a best answer is worth $10$ points.

A leaderboard, updated daily, ranks users based on the total number of points they collected.
%\ainote{Need to explain the levels here: how many, how are they defined, what are the consequences.}
%\ainote{Any related work that studies points, incentives, etc in YA?}
Users are split into seven levels based on their acquired points (e.g., 1-249 points: level 1, 250-999 points: level 2, ..., 25000+ points: level 7).
These levels are used to limit user actions, such as posting questions, answers, comments, follows, and votes: e.g., first level users can ask $5$ questions and provide $20$ answers in a day.

\emph{YA} requires its users to follow the Community Guidelines that forbid users to post spam, insults, or rants, and the Yahoo Terms of Service that limit harm to minors, harassment, privacy invasion, impersonation and misrepresentation, fraud and phishing. 
Users can flag content (questions, answers or comments) that violates the Community Guidelines and Terms of Service using the ``Report Abuse'' functionality.
%Figure~\ref{fig:abuse_report} shows how reporting is done on a content.
Users click on a flag sign embedded with the content and choose a reason between violation of the community guidelines and violation of the terms of service.
%~\cite{YahooAnswersCommunity} 
%Users can select between two reasons--- violation of the community guidelines (e.g., chat or rant, adult content, spam, insulting other members, etc.) and violation of the terms of service (e.g., harm to minors, violence or threats, harassment or privacy invasion, impersonation or misrepresentation, fraud or phishing, etc.).
Reported content is then verified by human inspectors before it is deleted from the platform.
%\ainote{when does content get removed? what about the high level users that are not checked by human inspectors? who are the human inspectors? etc. IK: This is a very internal question. We probably can't reveal more.
%Everyone has a trust level. If reporter and poster trust difference exceeds a threshold then the content immediately removed. Otherwise, the report is evaluated  by customer care.}
%NI: I don't think we can say much more in detail about these inspectors without talking about the trust score system which is not public.

\ignore{
\begin{figure}[htbp]
\centering
\includegraphics[height=4.3cm]{abuses/sample_abuse_reporting.pdf}
\caption{Abuse reporting on an answer. Users can click on the ``flag'' sign of the answer and report an abuse on the answer.}
\label{fig:abuse_report}
\end{figure}
}

Users in \emph{YA} can choose to follow other users, thus creating a follower-followee relationship used for information dissemination. 
The followee's actions (e.g., questions, answers, ratings, votes, best answer, awards) are automatically posted on the follower's newsfeed. 
%\ainote{is this e.g. or should it be i.e.? that is, if the list within parenthesis is complete, than it should be i.e.}
In addition, users can follow questions, in which case all responses are sent to the followers of that question.
% \ainote{``which is another information shared with that user's followers.'' Should this be maybe:  ``in which case all responses are sent to the followers of that question''}
%A user's followers can also learn whether his answer is selected the best or whether he is following a question.
%\ainote{so, there is a feature for following questions, as well? IK: yes, they can}

%follower-followee (FF) directed social graph.
%Users can follow each other using a  ``follow'' functionality from their homepage.
%If a user A is following another user B, then an edge from A to B ($A\rightarrow B$) is generated.
%In this case, A is a follower of B and B is a followee of A.

%-----------------------------------------------
\subsection{Dataset}
%-----------------------------------------------

\begin{figure}[htb]
\centering
\includegraphics[height=5cm]{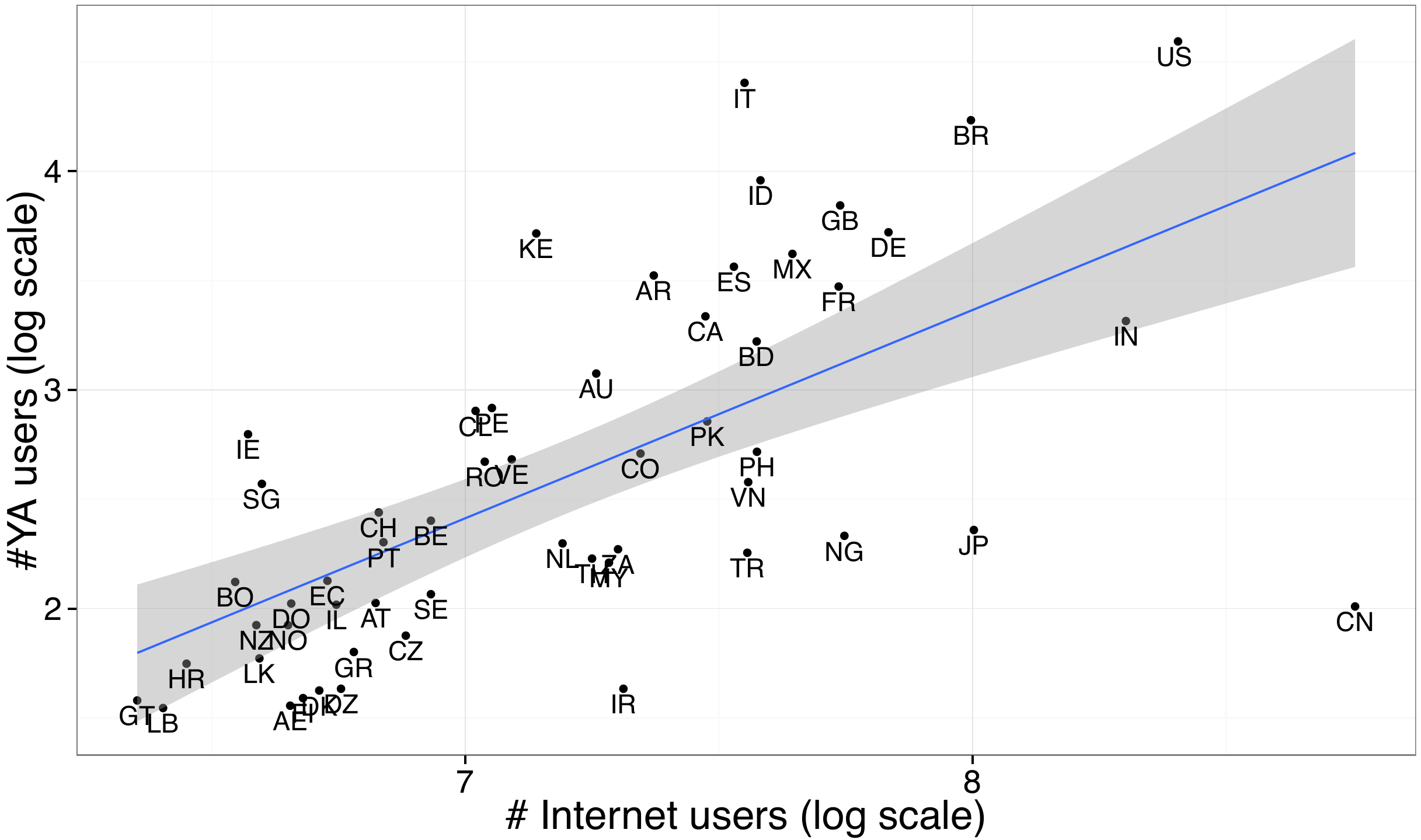}
\caption{The number of Internet users and YA users for 67 countries. The regression line and 95\% confidence interval area are also shown.
The countries are represented by a 2-letter country code based on ISO 3166.}
\label{fig:intYA}
\end{figure}

We studied a random sample of about $200$k users from \ya\ who were active between 2012 and 2013.
These users posted about $9$ million questions (about $45$ questions/user), 43 million answers (about $215$ answers/user), and 4.5 million abuse reports (about $23$ reports/user).
They are connected via $490$k  follower-followee relationships in a social network.
The indegree and outdegree distributions of the social network follow power-law distributions, with an exponential fitting parameter $\alpha$ of $1.83$ and $1.85$, respectively.
In our dataset, we have users from $67$ countries.
Figure~\ref{fig:intYA} shows the number of users in our dataset as a function of the number of Internet users taken from the World Bank.\footnote{http://www.data.worldbank.org/indicator/IT.NET.USER.P2}
We find a linear relationship between the number of users per country in our \ya\ dataset and the number of Internet users in the World Bank dataset for each country.
It means that the \ya\ users from our dataset are not skewed by country. 
Instead, they represent a sample of global Internet users.
%However, we don't want to consider those countries for analysis who are not represented well in the dataset, i.e., some of the countries in our dataset  have less than 100 users.
%One the other hand, we want to take enough number of countries which allows us to draw meaningful conclusions.
 
\begin{figure}[htb]
\centering
\includegraphics[height=5cm]{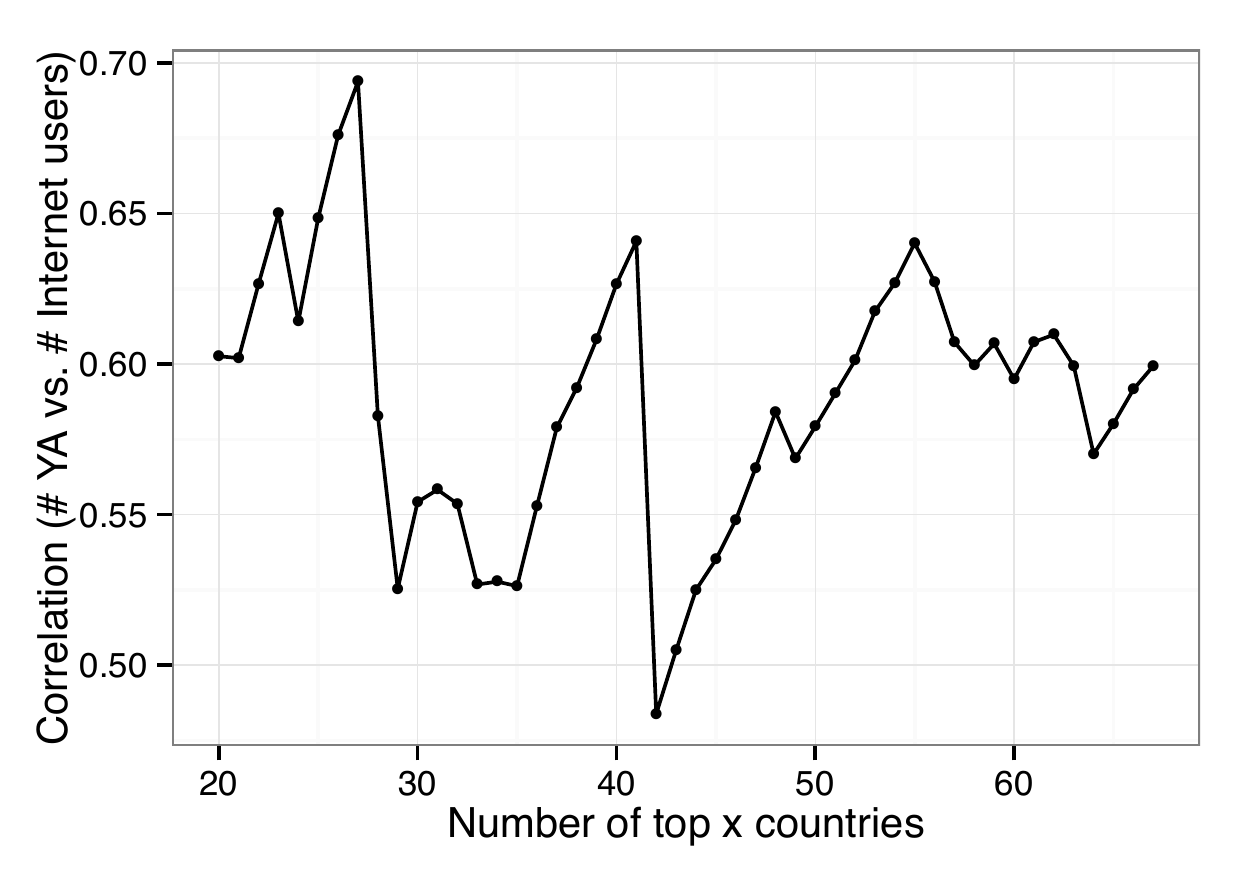}
\caption{Number of top countries based on the number of YA users and correlation with their number of Internet users. All correlations are statistically significant with $p$-value<0.05.}
\label{fig:corCountry}
\end{figure}

To investigate how sensitive this correlation is to the number of users per country, we computed the Pearson correlation between the number of \ya\ users in $x$ countries and their respective internet population.
The $x$ countries were ranked based on the number of \ya\ users found in the dataset, and $x$ was varied from top 20 to all 67 countries.
Figure~\ref{fig:corCountry} shows that there are several peaks in the correlation, but the values are high and between 0.5 and 0.7.
We select as a threshold the second highest correlation peak and thus included in the study $41$ countries which have at least 150 users per country.

%\section{Cultures in Yahoo Answers}\label{sec:culture}

%We analyze a number of cultural dimensions in Yahoo Answers proposed by Robert V. Levine and Geert Hofstede.
%Section~\ref{pace-of-life} presents Levine's Pace of Life and relates it to Yahoo Answers.
%Section~\ref{hofstede-culture} shows how three cultural dimensions---individualism, power distance and uncertainty avoidance---defined by  Geert Hofstede  are manifested in the ecosystem of Yahoo Answers.

\section{Levine's Pace of Life}\label{sec:pace-of-life}
In this section, we analyze Levine's Pace of Life cultural dimension in the context of \ya\ and show how it relates to user activities such as questioning, answering and reporting.
In his book~\cite{Levine2006Time}, psychologist Robert Levine defines Pace of Life as ``the flow or movement of time that people experience''. 
With the help of 19 experimenters, he collected and compared three indicators of the Pace of Life in 36 large cities from 31 countries around the world during a warm summer month between 1992 and 1995~\cite{levine1999pace}. The indicators are: 

\begin{itemize}
\item \textbf{Walking speed:} 
They measured walking speed of 35 men and 35 women over a distance of 60 feet in main downtown areas in each city.  Measurements were done during prime business hours after controlling a number of variables such as sidewalks, crowd, effects of socialization. 
They found significant differences in pedestrians walking speed---for example, pedestrians in Rio de Janeiro, Brazil walked only two-thirds as fast as pedestrians in Zurich, Switzerland.

\item \textbf{Postal speed:}
In each city, they measured the time it took postal workers to serve a standard request for stamps and considered this time as a proxy for work speed. 
They handed each clerk money and a note written in the local language requesting a common stamp. % was also given to them.
For example, in the United States, the clerk was handed a 5 dollar bill with a request for one 32-cent stamp. 
They found that overall Western Europe was the fastest to serve a standard request.

\item \textbf{Clock accuracy:}
To quantitatively measure time concerns, the researchers checked the clock accuracy of randomly selected 15 downtown banks in each city. 
The reported times were then compared to those reported by the telephone company, which was considered accurate.
\end{itemize}

Levine combined these three scores into a country-specific Pace of Life score and concluded that  ``the
Pace of Life was fastest in Japan and the countries of Western Europe and was slowest in economically undeveloped countries. The pace was significantly faster in colder climates, economically productive countries, and in individualistic cultures''~\cite{levine1999pace}.
%Faster places also tended to have higher rates of death from coronary heart disease, higher smoking rates, and greater subjective well-being''~\cite{levine1999pace}.

Intuitively, to cope with the rigid perception of time, people from the higher Pace of Life countries have to be planned and organized in their daily activities.
On the other hand, people from lower Pace of Life countries might allow some unstructured activities, as in those countries the expectation of following  the `right' time is more relaxed. 

%(e.g., \ainote{how else? why is worth mentioning? I think you mean something else. perhaps only 33 minutes and more as being late?} Levine's study found that students consider 33 minutes late as  `being late' in Brazil~\cite{levine1999pace}).

%We hope that these temporal fingerprints of the countries are not only evident in the offline world, but also are manifested in online space such as community question answering. \ainote{we hope?! why do we?}

Applying these findings to online communities such as CQA platforms, we expect that people from higher Pace of Life countries, such as the USA, will be less likely to ask or answer questions during busy hours of the day, e.g., office hours.
From these ideas, we  hypothesize the following in \ya:\\

\noindent
\textbf{[H1] \textit{Users from countries with a higher Pace of Life score show more temporally predictable activities.}} \\

To test this hypothesis, we calculate how probable a country's users are in asking, answering and reporting at different times of day and correlate that with that country's Pace of Life rank.
For example, if a user only asks or answers questions in the evening, he is temporally more predictable than a user who asks or answers in the morning, afternoon and night.
In a Twitter study~\cite{golder2011diurnal}, Golder and Macy also find diurnal mood rhythms in different cultures.

In order to calculate temporal predictability, we only consider working days, as weekends are less predictable. % (e.g., some people catch up on sleep).
More specifically, similar to~\cite{garcia2013cultural}, we divide the working day in five time intervals: morning (6:00 - 8:59), office time (9:00-17:59), evening (18:00-20:59), late night (21:00- 23:59), sleeping time (00:00 - 05:59).  
All the reported times are users' local time.
We use \emph{information entropy}~\cite{cover2006elements}, a measure of disorder, to calculate the temporal predictability.

%Then, to capture each user?s predictability, we compute the user?s entropy in those five intervals by: 
%Say we have C intervals; p(c) is the probability that a user posted (questions/questions/reports) in interval c. Then the entropy

%\begin{verbatim}-?_c?C??p(c)log p(c)?/log|C|\end{verbatim}

For a given activity (asking, answering, or reporting) and $C$ intervals, we can compute $p(c)$, the probability of an activity belonging to interval $c$. 
We measure the normalized entropy  for user $u$  for all activities as:

\begin{equation}
Entropy_{u}=  \frac{-\sum_{c \in C}p(c)log(p(c))}{|logC|}
\end{equation}

We calculate users' normalized entropies for all their questions, answers and abuse reports and refer to them as \emph{question}, \emph{answer} and \emph{report entropy}, respectively.
In our dataset, each country has on average $134$k questions, $642$k answers and $67$k abuse reports.
Normalized entropy ranges from 0 to 1. 
A normalized question entropy close to 0 indicates that most of the questions the user asked are within one time interval of the day, whereas the closer to 1, the more likely is that the user asked questions during all intervals. 
%Finally, for a country, we report the geometric mean of its users' question, answer and report entropies.
Finally, the question/answer/report entropy  for a country $c$, $Entropy_{q/a/r,c}$, is defined as the geometric mean  of all $Entropy_{q/a/r,u}$  computed for the users of that country:
\begin{equation}
Entropy_{q/a/r,c} =\Bigl( \prod \limits_{ u \in U_c} Entropy_{q/a/r,u}\Bigr)^{\frac {1}{|U_c|}}
\end{equation}
where $U_c$ is the set of users in country $c$. We use geometric mean to account for the skewed distribution of the entropy scores, something that the regular arithmetic mean cannot handle.
%The use of arithmetic mean as an alternate doesn't consider this skewness  and the average is dominated by a range of numbers (e.g., too high or low).

\begin{figure}[ht]
\centering
\includegraphics[height=5cm]{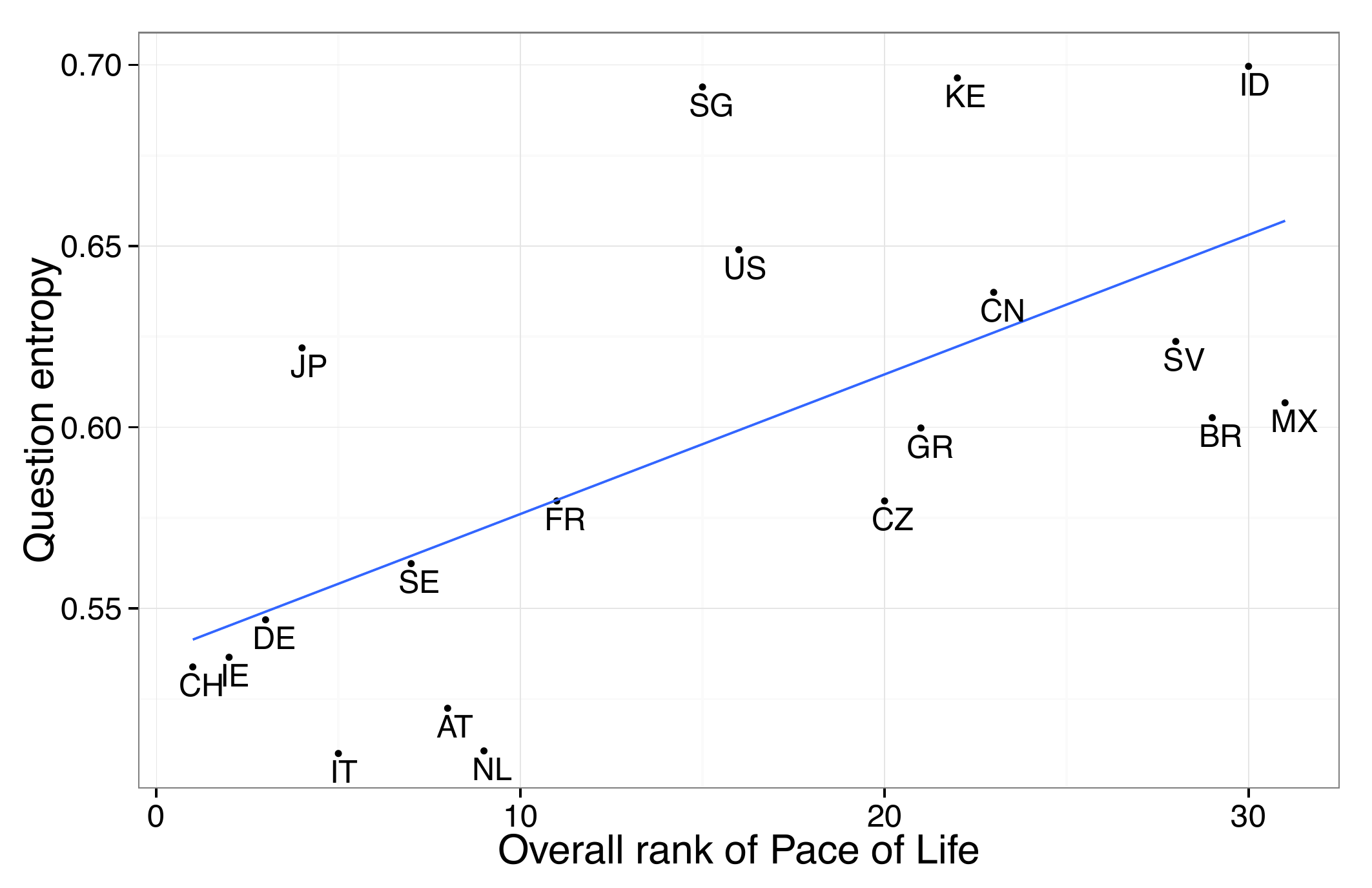}
\caption{Pace of Life overall rank vs. average question entropy per country.  Countries shown are the ones in our dataset for which a Pace of Life rank has been published. Countries are ranked in decreasing order of their Pace of Life value. A regression line is also shown. }
\label{fig:question-entropy}
\end{figure}

\begin{figure}[ht]
\centering
\includegraphics[height=5cm]{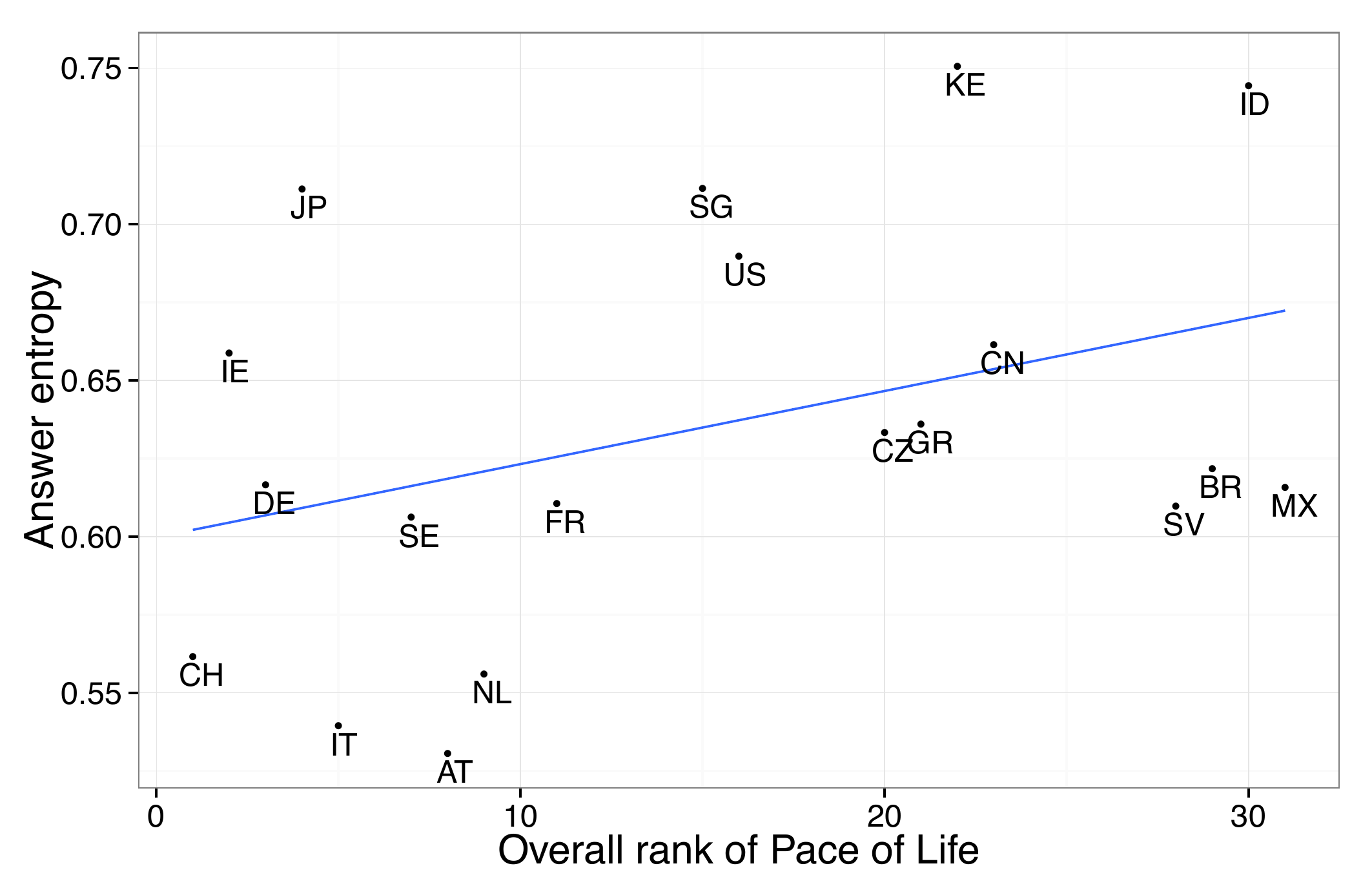}
\caption{Pace of Life overall rank vs. average answer entropy per country.  Countries shown are the ones in our dataset for which a Pace of Life rank has been published. Countries are ranked in decreasing order of their Pace of Life value. A regression line is also shown.}
\label{fig:answer-entropy}
\end{figure}

Table~\ref{entropy-paceoflife} shows Pearson correlations between question, answer, report entropy and Pace of Life ranks, where lower ranks mean higher Pace of Life.
For both questions and answers, the overall Pace of Life ranks have positive correlations with question and answer entropy with $r=0.67$ and $r=0.37$, respectively.
These positive relationships are seen in the Figures~\ref{fig:question-entropy} and ~\ref{fig:answer-entropy}. 
We find positive correlations between walking speed rank, post office service time rank, and clock accuracy time rank with question entropy  with $r=0.48$, $r=0.60$ and $r=0.48$, respectively.
For answers, we find positive correlations between post office service time rank, and clock accuracy time rank with entropy with $r=0.38$ and $r=0.29$, respectively.
However, we do not find any statistically significant relationships between report entropy and Pace of Life ranks.

These results confirm that users from countries with a higher Pace of Life score show more temporally predictable asking and answering behavior in \ya.
%%Reporting 
%Report is not significant, may be because reporting is sparse, only few people participate and clearly their patten is not normal to asking and %answering (relevant the reporting distributions here).

%The entropy (0<=entropy<=1) reflects the (un) predictability of a user posting across all intervals.  Following are the correlations between countries entropies for question/answers/reports and Pace of Life rank (higher ranks mean low Pace of Life). Only question are significant, meaning that high Pace of Life counties are temporarily predictable in terms of when people are asking question and answering. 

\begin{table}[ht]
\centering
%\scalebox{0.8}
%{
\begin{tabular}{|l|l|l|l|l}
\hline
%\textbf{Name} & Value \\
%\hline
%\hline
&\multicolumn{3}{|c|}{Entropy}\\
\hline
\textbf{Pace of Life} 	&	\textbf{Question}	& \textbf{Answer}	& \textbf{Report}\\
\hline
% \textbf{`Suspended' (+ve class)} &100,162&Q: 77,626, A: 78,129\\
 %\hline
 %\textbf{`Fair' (-ve class)} &99,838& Q: 172,374, A: 171,871\\
 %\hline 
Overall			&	0.67*** 	& 	0.37*		& 	0.18\\
\hline
Walking speed		&	0.48**	& 	0.18		&	0.06\\
\hline
Post office			&	0.60**	&	0.38*		&	0.19\\
\hline
Clock accuracy		&	0.48**	&	0.29*		&	0.21\\
\hline
\end{tabular}
%}
\caption{Pearson correlations between question, answer, report entropy and Pace of Life rank. Lower ranks mean higher Pace of Life. $p$-values are indicated as: $p$<0.005(***), $p$<0.05 (**), $p$<0.1 (*).}
\label{entropy-paceoflife}
\end{table}

\section{Hofstede's Cultural Dimensions}\label{sec:hofstede-culture}

%%In their book~\cite{hoftede2010cultures}, Dutch social psychologist Geert Hofstede and his colleagues describe culture as a set of dimensions that separates one group of people from others.  
%%The dimensions contribute observable behavioral differences. 
%%For example, in USA laughter is related to happiness, or at least it has a positive connotation, whereas in Japan laughter is often considered a sign of confusion, insecurity and embarrassment. 
\ignore{
\ainote{This looks like fluff to me. Plenty. }
Hofstede and his colleagues create a pyramid structure (Figure ~\ref{fig:pyramid}), describing  culture as a collective behavior, and placing it inside personality and human nature.
Personality is inherited and learned characteristics and specific to individuals, as opposed to a culture that is a collective behavior learned in different societies. 
On the other hand, human nature is a universal inherited characteristic of humans.

\begin{figure}[ht]
\centering
\includegraphics[height=5.2cm]{pyramid.pdf}
\caption{The culture pyramid~\cite{hoftede2010cultures}.}
\label{fig:pyramid}
\end{figure}
}

In this section, we analyze a number of cultural dimensions in \ya\ proposed by Geert Hofstede.
We show how three cultural dimensions defined by Hofstede---individualism, power distance and uncertainty avoidance are manifested in the ecosystem of \ya.

Hofstede's cultural dimensions theory is a framework for analyzing cultural variability. 
In his original model~\cite{hofstede1983national}, Hofstede proposed four primary dimensions by surveying in the 1960s and 1970s a large number of IBM employees from 40 countries: power distance (PDI), individualism (IDV), uncertainty avoidance (UAI) and masculinity (MAS).  
Later~\cite{hoftede2010cultures}, he added two more dimensions: long-term orientation (LTO) and indulgence versus restraint (IVR).  
Three of the dimensions, individualism, power distance, and uncertainty avoidance, have been used in a number of recent studies of online behavior~\cite{garcia2014twitter,park2014cross,reinecke2013doodle}.
We also use these three  Hofstede's cultural dimensions and relate them to a number of hypotheses in the context of \ya.

%\vspace{-3mm}
\subsection{Individualism (IDV)}

Individualism is the extent to which an individual is integrated into a group. In individualistic societies (high IDV) such as the USA and England, personal achievements and individual rights are emphasized; an individual is expected to take care of only himself and his immediate family. 
In collectivist countries such as those of India, China, and Japan, individuals are expected to place the family and group goals above those of self. %a tightly-knit framework in society exists which expects an individual to work for family and group goals above his needs and desires.
In this work, we investigate how individualism is related to users' contribution, (un)ethical behavior  and privacy settings in \emph{YA}. 
\\

\textbf{Individualism and contribution.}
The usage of the Internet takes time from a number of daily activities including face-to-face socialization.
In collectivist countries, people are expected to give a fair amount of time on sociability, hence traditionally they seem to spend less time on the Internet compared to the people from the individualistic cultures~\cite{de2013global}.
In \ya\ we expect that users from individualistic countries spend more time online, hence they can provide more answers and eventually they can contribute more to the community than their direct benefits from the community.
We hypothesize the following:\\

\noindent
\textbf{[H2] \textit{ Users from countries with higher individualism index provide more answers.}} 
\\
\textbf{[H3] \textit{ Users from countries with higher individualism index contribute more to the community than what they  take away from the community.}} \\

\begin{figure}[ht]
\centering
\includegraphics[height=5cm]{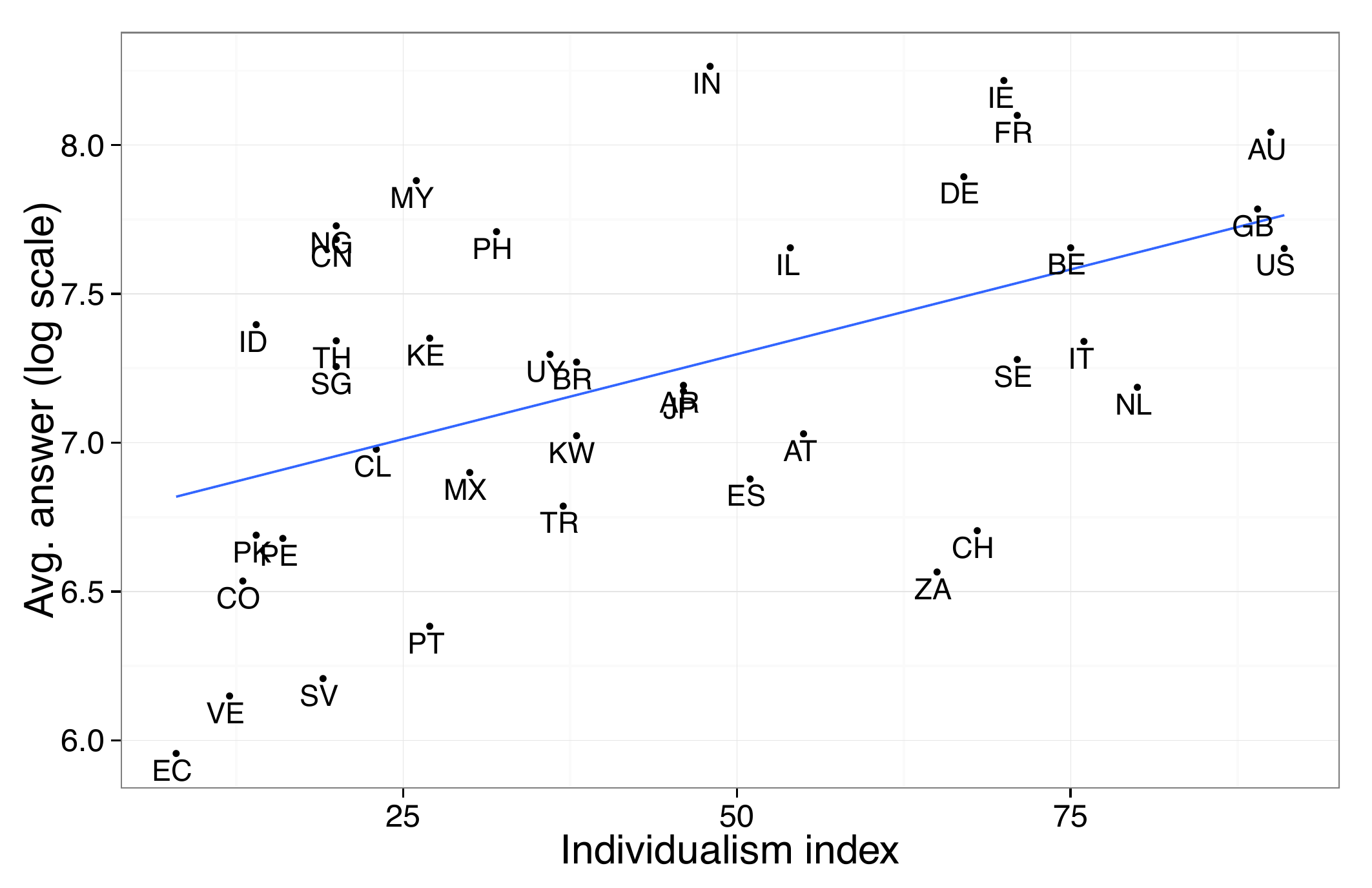}
\caption{Individualism index vs. the average number of answers posted by users per country.  A regression line is also shown.}
\label{fig:indVsans}
\end{figure}

We correlate the geometric mean of the number of answers posted by the users from a country with that country's individualism index (a higher score means higher individuality). 
We use geometric mean as an average because of the skewed distributions of the number of answers.
In the calculation of the geometric mean, we exclude the users who have not provided any answers.
We observe a positive correlation, shown in Figure~\ref{fig:indVsans}, with $r= 0.46, p<0.005$. 
This means that, on average, users from individualistic countries provide more answers.

To quantify users' contribution compared to their take away, we compute \emph{yielding scores} of the users.
The \emph{yielding score}  of a user is simply  a difference between his contribution and his take away.
For yielding  scores, we consider  \ya's point system, which awards two points for an answer, ten points for a best answer, and penalizes five points for a question:

\begin{equation}
\begin{split}
 \textrm{Yielding}_{u} &= f(contribution)-f(take away)\\
			&=2.0*A_u +10.0*BA_u-5.0*Q_u
\end{split}
\end{equation}
where $Q_u$ is the number of questions posted by user $u$, $A_u$ is the number of answers posted by $u$, and $BA_u$ is the number of best answers posted by $u$.

\begin{figure}[ht]
\centering
\includegraphics[height=5cm]{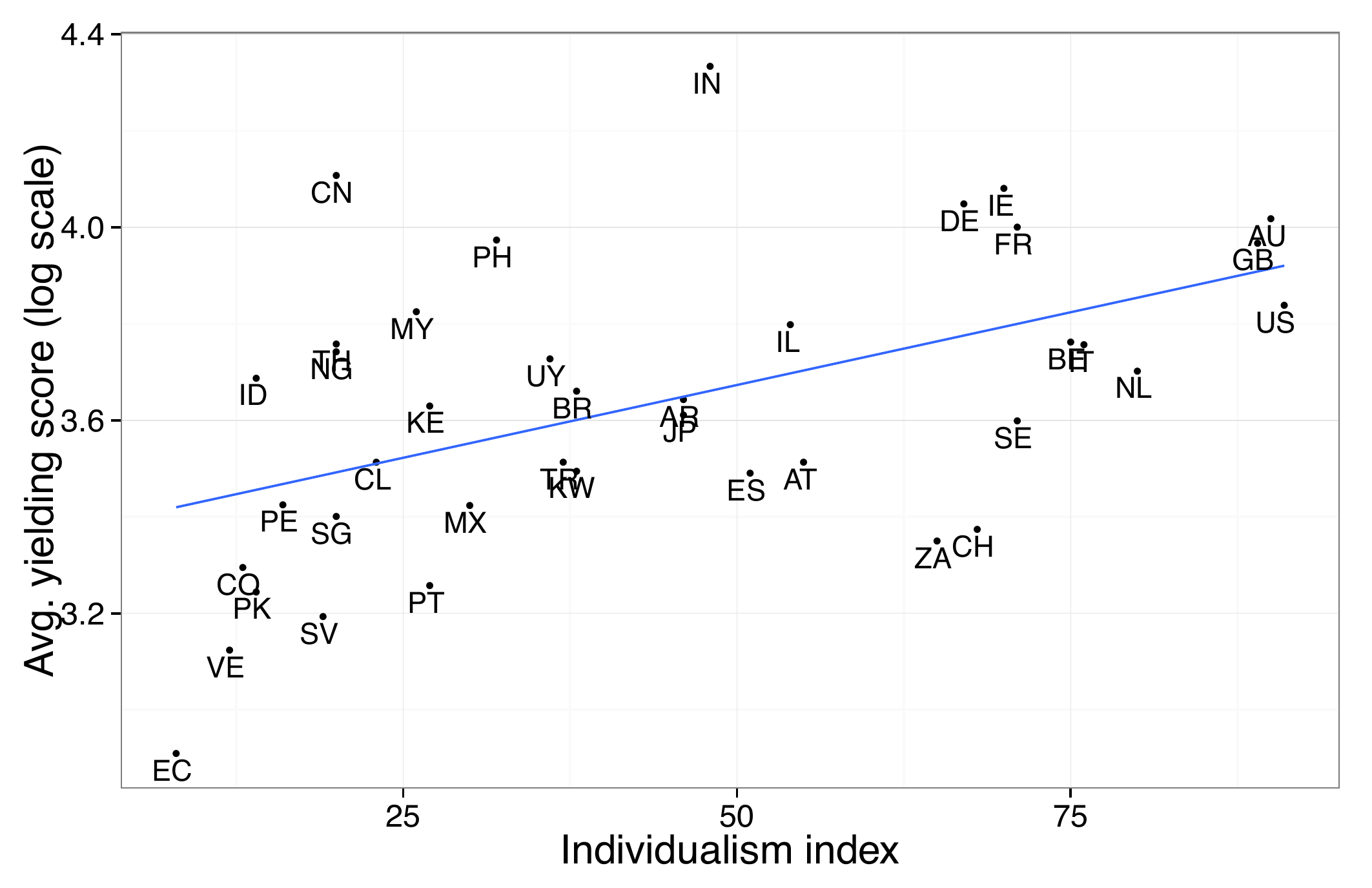}
\caption{Individualism index vs. yielding score per country. A regression line is also shown.}
\label{fig:ind_alt}
\end{figure}

Finally,  a country's yielding score $Yielding_c$ is defined as the geometric mean  of all $Yielding_u$  computed for the users of each country $c$:
\begin{equation}
Yielding_c =\Bigl( \prod \limits_{ u \in U_c} Yielding_{u}\Bigr)^{\frac {1}{|U_c|}}
\end{equation}

where $U_c$ is the set of users in country $c$ and we take only those users having yielding scores more than zero.
We correlate a country's geometric mean of the yielding score  with the country's individualism index and we obtain a positive correlation (Figure~\ref{fig:ind_alt}) with $r=0.37, p<0.05$.
%shows the positive relations between altruistic scores  and country individuality index.
This result suggests that  the more  individualistic a country is, the more its users contribute to \ya\ than what they take away from the community.\\

There might be multiple explanations about why users from individualistic countries contribute more to the community as reflected by hypotheses H2 and H3.
One explanation is that individualistic cultures have a more favorable collaborative environment~\cite{sosik2002work}, so individuals feel the urge to contribute to the community.
Another explanation could be that users from individualistic cultures simply want more points than collectivist cultures.
As points are awarded for contribution (e.g., an answer earns two points) and participation (e.g., each login earns one point), users might be tempted to contribute more.
In fact, we obtain a positive and significant correlation ($r=0.42, p<0.01$) between a country's points (calculated as geometric mean of the country's user points) and its individualism index. 
Finally, there might be other confounding factors (e.g., internet penetration) that affect the contribution of a country's users on the platform.
Thus, it is difficult to confirm whether the users' behavioral differences on contribution are due to their cultural differences.

\textbf{Individualism and (un)ethical behavior.}
The degree to which a culture is collectivist or individualistic has an implication on its users' online (un)ethical behavior.
For example, the more individualistic (less collectivistic) a culture, the lower the rate of software piracy~\cite{husted2000impact} and online music piracy~\cite{ki2006exploring}.
Personal rights are paramount in individualistic cultures, where people do not feel obligated to engage in group cooperation that involves conspiracy.
Group cooperation and conspiracy are two key elements for the real world unethical behaviors such as corruption~\cite{park2003determinants}.
Triandis et al.~\cite{triandis2001culture} used Hofstede's individualism index and found that the countries with higher collectivist scores show the most corruption.

Based on this online and offline user unethical behavior that is influenced by culture, our intuition is that we could observe a similar trend in \ya.
%\ainote{is the following already described in the dataset? if so, we should not repeat it, i think}
%\ainote{ need to rewrite this:
In CQA platforms, the expectation is that users would provide helpful answers to posted questions.
As such, users are required to follow the Community Guidelines and the Yahoo Terms of Service while answering.
When users post bad answers, community members flag them.
Later, human moderators check whether these flags are applied correctly or not.
We expect that the more collective a culture is, the more probable the answers from its users will be flagged as abusive.
%} 
%\ainote{you're saying that the expectation is that incorrect flagging is unethical -- vindicative, for example.}
Formally, we hypothesize that:\\

\noindent
\textbf{[H4] \textit{Users from more collective (less individualistic) cultures have higher probability to violate CQA norms.}} \\

To this end, for each user $u$, we first calculate  $p_u$, the probability that his answers violate community norms (and thus are correctly flagged by other users): %, i.e., correctly flagged by the users:
\begin{equation}
\centering
\begin{split}
 \textrm{p}_u=  \frac{ \textrm{\# correctly flagged answers from u}}{\textrm{\# total answers from u }}
 \end{split}
 \label{eq:flagged_probability}
\end{equation} 

Finally, $P_c$, the geometric average of all $p_u$ probabilities computed for each country $c$:
\begin{equation}
P_c =\Bigl( \prod \limits_{ u \in U_c} p_{u}\Bigr)^{\frac {1}{|U_c|}}
\end{equation}
where $U_c$ is the set of users in country $c$. 

\begin{figure}[htbp]
\centering
\includegraphics[height=4.7cm]{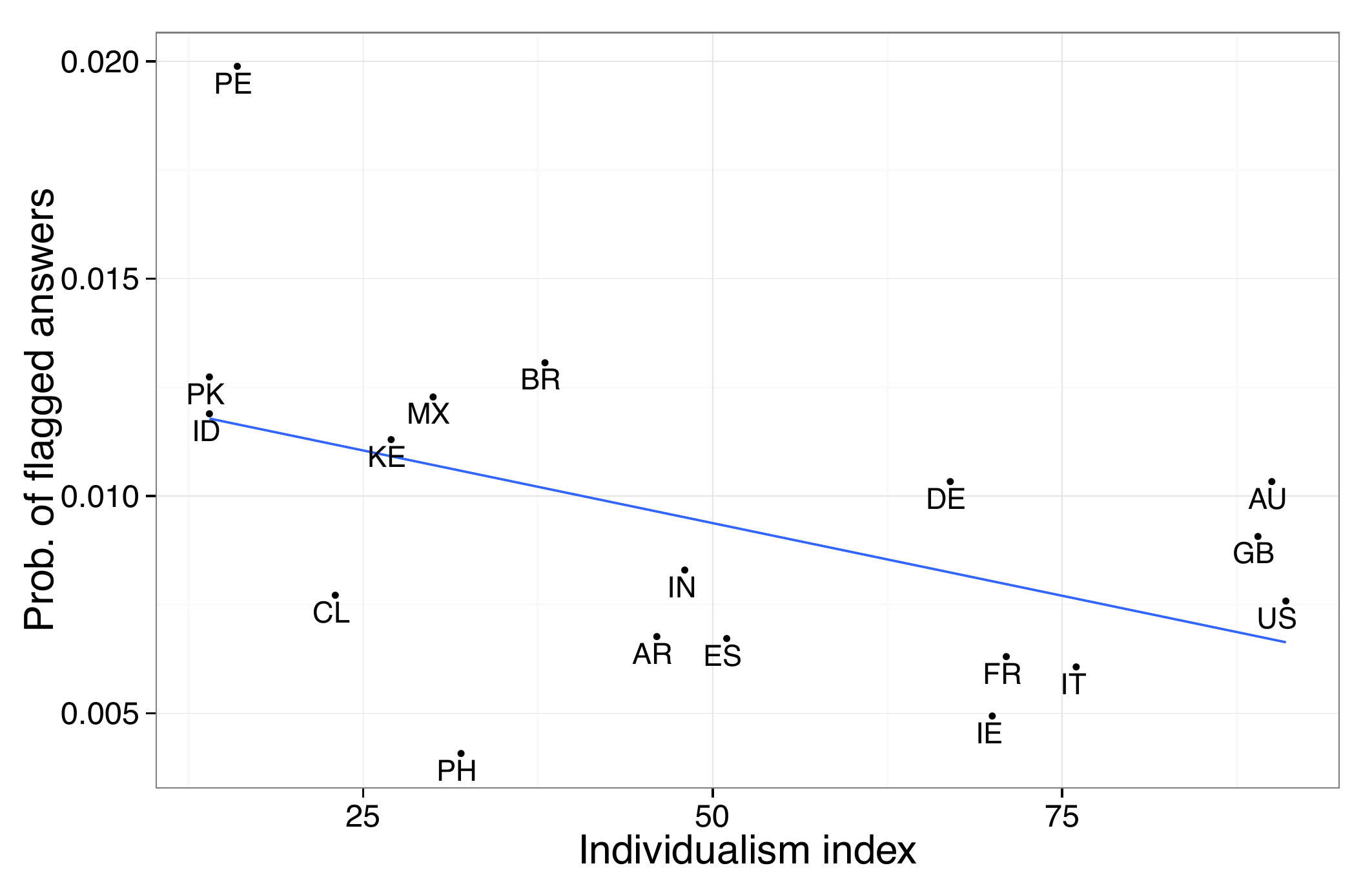}
\caption{Individualism index vs. the probability that an answer from a country is correctly flagged.  A regression line is also shown. }
\label{fig:ind_flagg}
\end{figure}

The Pearson correlation $r=-0.48, p<0.05$  shows that the probability of abuses in answers provided by the users from a particular country is negatively correlated with that country's individualism index. 
%\ainote{is this what you meant?}
%that an answer from a country is correctly flagged  and its individualism index is negatively correlated.
%The figure hints that the relationship is negative.
Figure~\ref{fig:ind_flagg} indeed shows that the probability decreases with an increasing individualism index, meaning that if an answer comes from an individualistic country, it is less probable to violate community rules.

\textbf{Individualism and privacy concerns.}
Although online privacy concerns are global, the extent to which people perceive these concerns as real varies across cultures.
For example, in the United States, privacy is a basic human right, endorsed by the American Bill of Rights, while Asian countries show little or no recognition on privacy in their legal systems~\cite{de2013global}.
%People in collectivistic cultures tend to accept the intrusion of groups and organizations in their private life. %, as opposed to individualistic cultures, where people place more value on private life.
%As such, researchers have proposed that users from individualistic countries would show higher levels of online privacy concerns~\cite{liu2004american,milberg1995values}.
%In fact, 
A survey of $1261$ Internet users from five big cities---Bangalore, Seoul, Singapore, Sydney and New York---shows that Internet users from individualistic cultures are more concerned about privacy than those in collective cultures~\cite{cho2009multinational}.
We expect that  a similar trend also exists in CQA platforms.
We hypothesize that:\\

\noindent
\textbf{[H5] \textit{Users from higher individualism index countries exhibit higher level of concern about their privacy.}} \\

We use the modifications of the privacy settings on users' \ya\ accounts as a proxy of privacy concern.
In \ya, privacy settings are typically available for users to personalize for content (questions or answers) and follower-followee network.
Intuitively, privacy-concerned users would take the opportunity to change the default privacy settings.
So, we consider the fraction of public privacy profiles in a country to draw a conclusion on how concerned its users are about their privacy.
However, the default privacy in YA is public.
It might be possible that many of the users in the public group are dormant: users who signed up, asked and answered some questions, and disappeared quickly. 
These users might skew the results of our study, thus, we only consider active users from our dataset--- users who have asked and answered more than 10 questions during our observation interval.
These active users are about 79\% of our dataset. 
We note that our conclusions remain the same if we consider more active users by filtering users who have asked and answered more than 20 questions. 

\begin{figure}[ht]
\centering
\includegraphics[height=4.7cm]{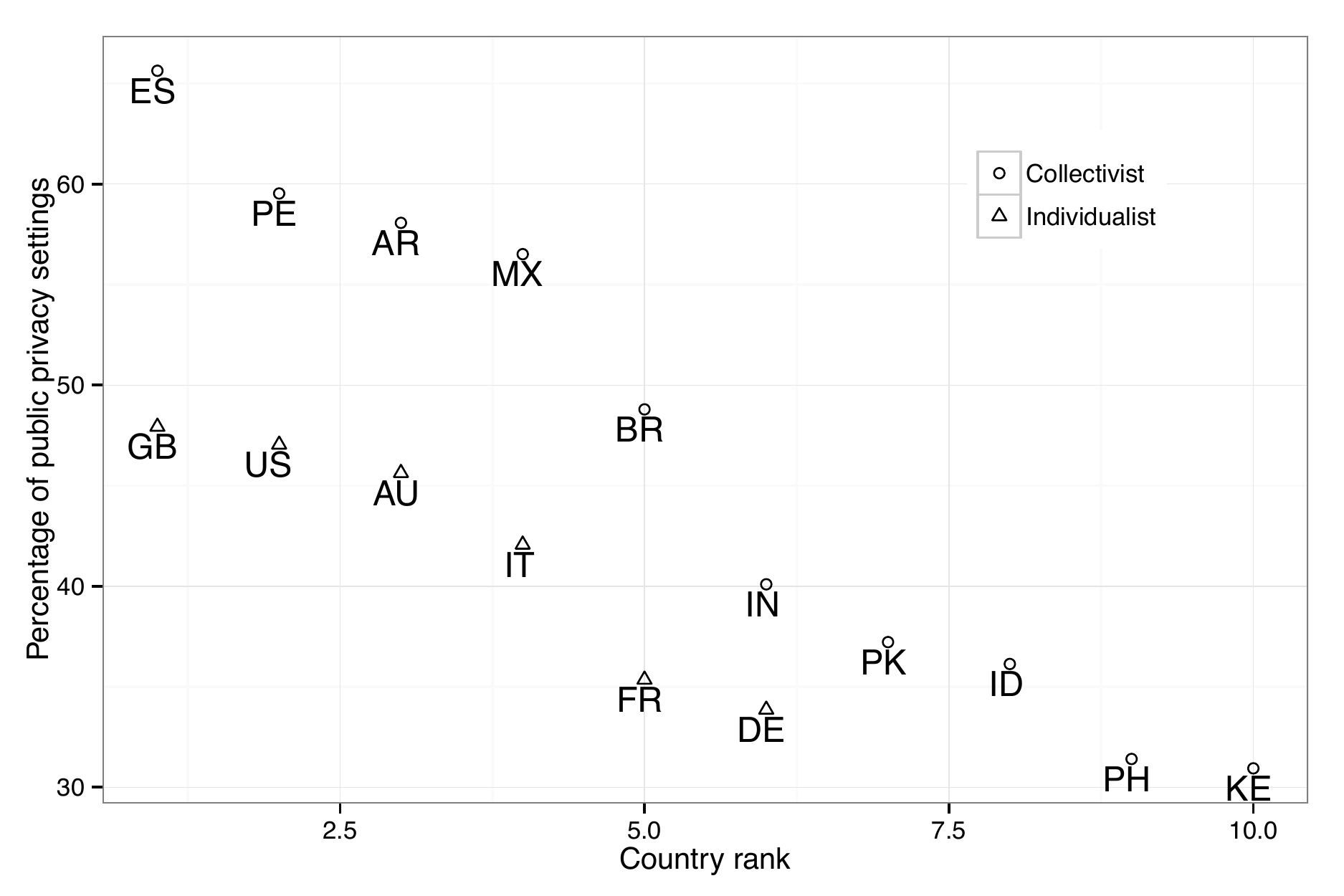}
\caption{Percentage of public privacy settings vs. rank of collectivist and individualistic countries, respectively. Country ranks are based on the percentage of public privacy settings and they are separately done for collectivist and individualistic countries.}
\label{fig:ind_privacy}
\end{figure}

Based on Hofstede's Individualism index, the Hofstede Centre\footnote{http://geert-hofstede.com/countries.html} has tagged  countries as individualistic or collectivist.
In our study, we use this classification. 
Figure~\ref{fig:ind_privacy} shows the percentage of user profiles with public privacy settings in a country, as function of the country's ranking in the collectivist and individualistic class.
The figure shows that, on average, collectivist countries have a higher percentage of public profiles: collectivist countries such as Spain, Peru, Argentina, and Mexico have higher percentage of public profiles than individualistic countries such as United Kingdom, United States, Australia or Italy.

\subsection{Power Distance Index (PDI)}
PDI is the extent to which the less powerful members of an organization or society expect and accept that power is distributed unequally. 
This dimension sheds light on how a society handles inequalities among its members. 
In countries with high PDI, such as countries from Latin, Asian, African and Arab world, everybody has a place in the social hierarchy and people accept the situation without questioning it.  
However, in Anglo and Germanic countries, which are low power distance countries, people seek distribution of power and ask for justifications of power inequality.

PDI essentially measures the distribution of  wealth and power between people in a country or culture.
In \ya, we can use the indegree (number of followers) as a proxy of wealth and power.
For example, the larger the number of followers users have, the larger an audience they have for direct communication.
Higher indegree users are also found to be more central (thus more retained~\cite{imrul2014blogretentionSocialCom}) across a number of network centrality metrics~\cite{imrul12socinfo}.
Moreover, these users' questions are forwarded to more users, hence more likely to be getting an answer.
A study~\cite{Kayes2015WWW} on \ya\ shows that users receive more answers from close neighborhoods.
Given the high number of questions that remain unanswered (42\% in \ya\ reported by a study~\cite{richardson2011supporting}) in CQA platforms, bringing answers not only shows a user's potential capability, but also makes the platform mature and informative.  
Taking ideas from the unequal distribution of  wealth and power in higher power distance countries, we expect that in \ya, users from those countries also have inequality in their indegrees. 
Garcia et al.~\cite{garcia2014twitter} have found similar indegree inequality  in Twitter.
We hypothesize the following in \ya:\\

\noindent
\textbf{[H6] \textit{Users from higher power distance countries show a larger indegree imbalance in follow relationships.}} \\

%For Yahoo Answers, we can say (based on the Twitter paper)
%-In countries comfortable with Power Distance, a pair of users who engage in any type of relationship is likely to show indegree imbalance.
We correlate countries power distance index (higher index means power distance is high) with their users' indegree imbalance.
A user's indegree imbalance is calculated as the difference between her friends' average indegree  and her indegree.
Finally, a country's indegree imbalance is the geometric mean of the indegree imbalance of its users.

For all countries, except Panama and Philippines, we obtained a positive indegree imbalance, meaning that for those countries, on average, a user's contacts have more contacts than the user.
This supports a well-known  hypothesis \emph{friendship paradox} in sociology.
The friendship paradox states that your friends have on average more friends than you have,  however, most people think that they have more friends than their friends have~\cite{feld1991your}.
It has been shown that the paradox holds for both Twitter~\cite{hodas2013friendship} and Facebook~\cite{ugander2011anatomy}.
Now we also show it for \ya.

\begin{figure}[ht]
\centering
\includegraphics[height=5cm]{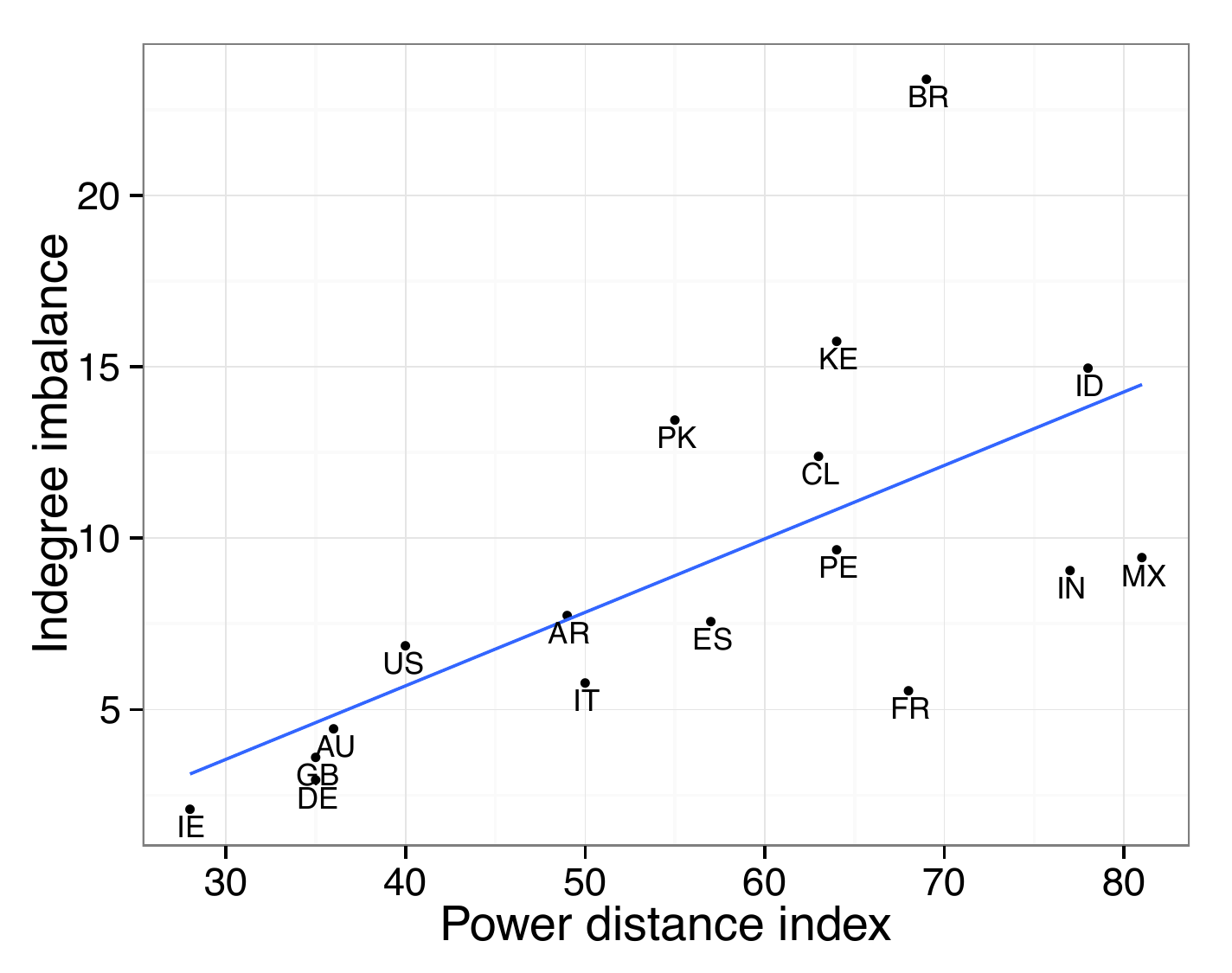}
\caption{Power distance index vs. indegree imbalance.  A regression line is also shown.}
\label{fig:pwi}
\end{figure}

Figure~\ref{fig:pwi} shows the relation between PDI and indegree imbalance (excluding Panama and Philippines).
The figure indeed shows a positive correlation.
We obtained a positive correlation $r=0.65, p<0.005$ between indegree imbalance and PDI for all countries (including Panama and Philippines). 
This supports the hypothesis that users from countries with higher PDI are more comfortable with indegree imbalance.

\subsection{Uncertainty Avoidance Index (UAI)}
UAI is the extent to which people feel uncomfortable with uncertainty and ambiguity. 
Individuals from countries exhibiting strong UAI tend to minimize uncertainty and ambiguity by careful planning, and enforcing rules and regulations. 
On the other hand, low uncertainty avoidance cultures maintain a more relaxed attitude in unstructured situations.

For example,  Switzerland has  a reasonably high uncertainty avoidance index (58) compared to countries such as Singapore (8) and  Sweden (29). 
%Swiss people  plan everything carefully and are not comfortable on uncertainty. 
In fact,  an online scheduling behavior study~\cite{reinecke2013doodle}  on Doodle (http://doodle.com/) shows that Switzerland and Germany have a high advance planning time of 28 days. 
In YA, our related hypothesis is:\\

\noindent
\textbf{[H7] \textit{Users from countries with higher uncertainty avoidance index exhibit more temporally predictable activities.}}\\

%For Yahoo Answers, we can say 
%-	The activities (e.g., questions, answers, abuse reports) of users in countries with higher uncertainty avoidance are also temporarily predictable

\begin{figure}[ht]
\centering
\includegraphics[height=4.7cm]{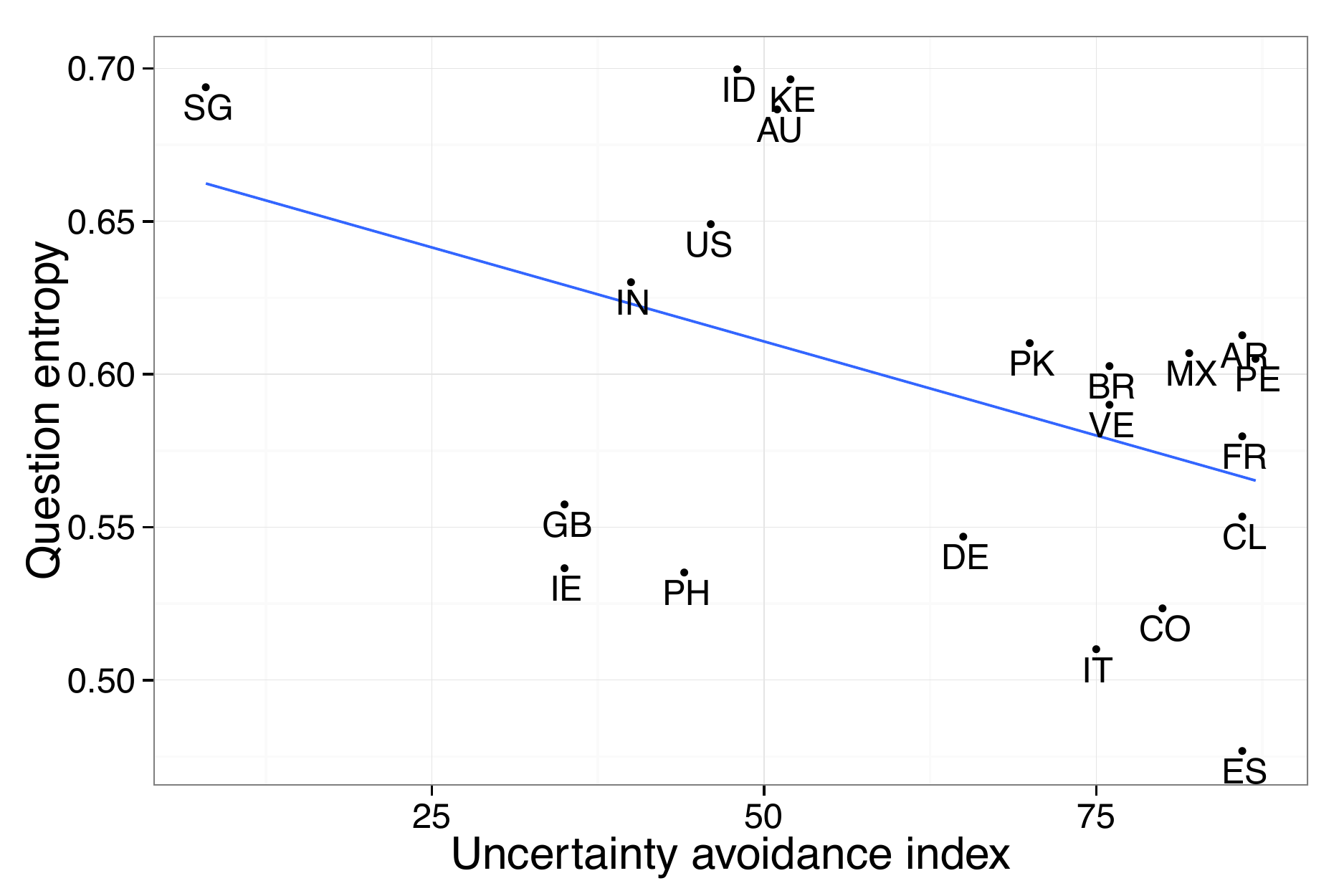}
\caption{Question entropy vs. uncertainty avoidance index. Only countries having more than 300 users are plotted. A regression line is also shown.}
\label{fig:qe-uai}
\end{figure}

\begin{figure}[ht]
\centering
\includegraphics[height=4.7cm]{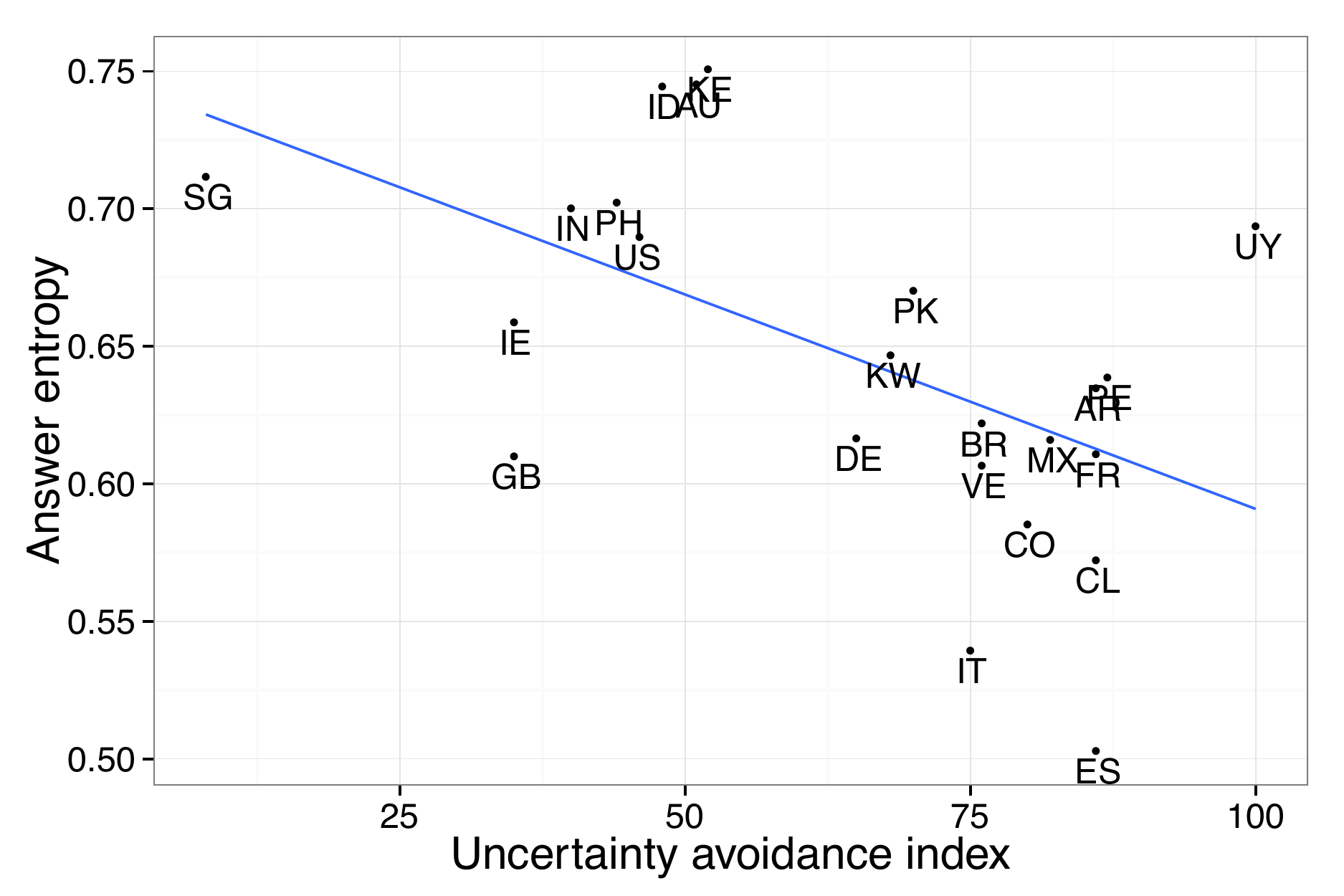}
\caption{Answer entropy vs. uncertainty avoidance index. Only countries having more than 300 users are plotted. A regression line is also shown.}
\label{fig:ae-uai}
\end{figure}

\begin{figure}[ht]
\centering
\includegraphics[height=4.7cm]{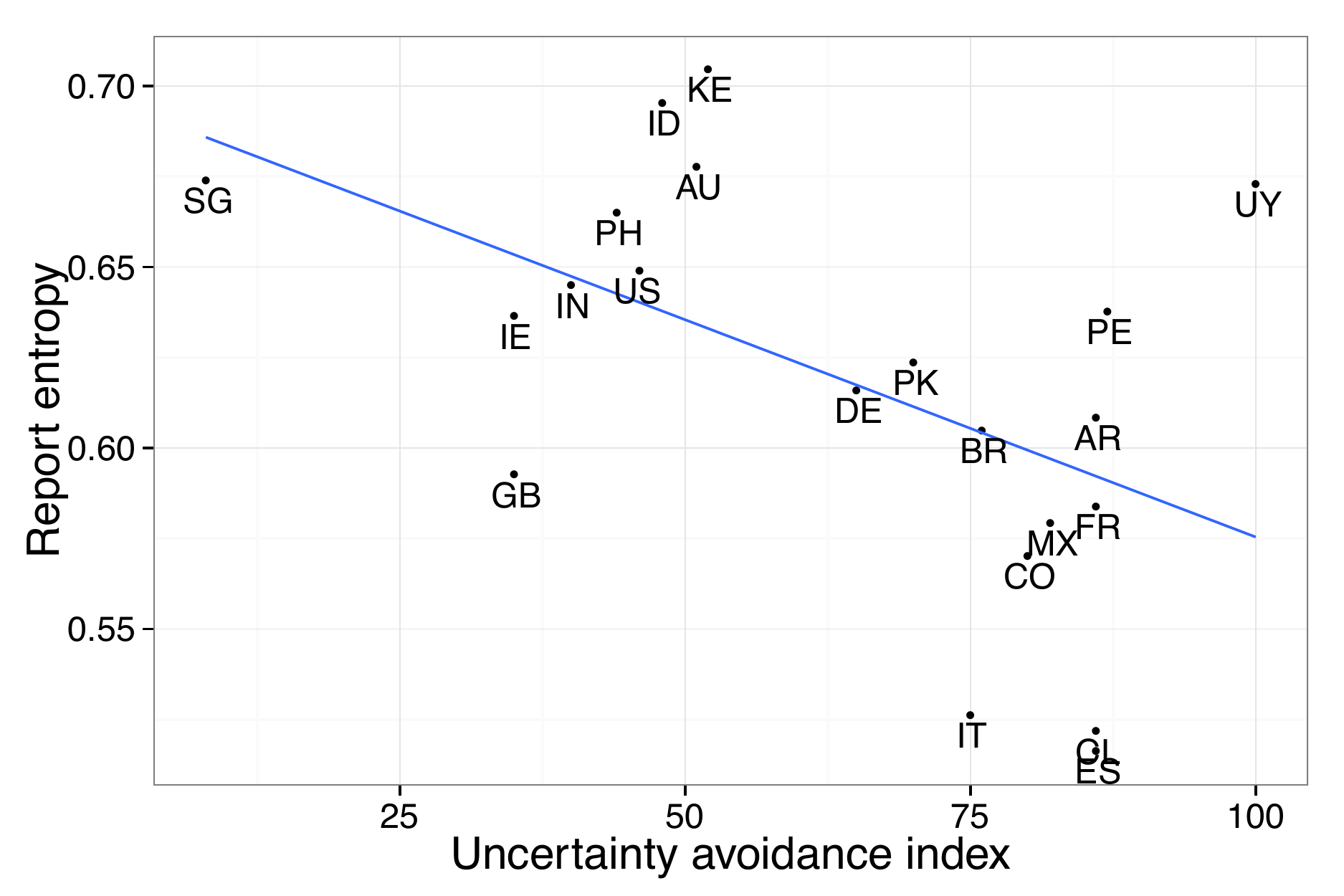}
\caption{Report entropy vs. uncertainty avoidance index.  Only countries having more than 300 users are plotted. A regression line is also shown.}
\label{fig:re-uai}
\end{figure}

Figures~\ref{fig:qe-uai},~\ref{fig:ae-uai},~\ref{fig:re-uai} show the relationship between question, answer and abuse report entropy vs. uncertainty avoidance index, respectively.
Note that a higher UAI means lower uncertainty and ambiguity.
 The negative relations in the figures indicate that users from countries with higher uncertainty avoidance index tend to have lower question, answer and abuse report entropies, thus they are more temporarily predictable.
% Table~\ref{correlation-uai} shows the Pearson correlation  between uncertainty avoidance index and entropies.
 All the entropies have negative relation to uncertainty avoidance index: $r=-0.43$ for questions, $r=-0.55$ for answers, and $r=-0.51$ for abuse reports. 
 All correlation values are statistically significant with $p<0.05$.

\section{Summary and Discussion}
\label{sec:discussion}

\ignore{
This work builds an understanding of the culture in Yahoo Answers.
%Using Hofstede's theory of cultural dimensions and Levine's Pace of Life,
We study users' behavioral patterns such as temporal predictability of activities, engagement through  contribution, and privacy concerns in Yahoo Answers in comparison with a number of dimensions (Pace of Life, Individualism, Uncertainty Avoidance and Power Distance) that vary across cultures. % and associate them with 
We discuss some implications of this work in the following.
}

Observing the global spread of  information and communication technologies, researchers sometimes predicted that the online world would be converging  into a ``one-world culture''~\cite{levitt2002globalization}.
%It seems, this vision is an illusion, thanks to the influence of so many cultural (as well as non-cultural) factors.
With the advent of the large-scale online behavioral datasets in the past decade from online  platforms like Twitter, Facebook and Foursquare, researchers showed that the Internet does not have a homogeneous culture.
Instead, country-specific cultural variations do exist.
We showed the same non-homogeneity, but in a very different online context---community question answering.

%\ainote{paragraph here summarizing the work and the results}

In this work, we analyzed about 200 thousand sampled Yahoo Answers users from 67 countries.
We studied users' behavioral patterns such as temporal predictability of activities, engagement, (un)ethical behavior, privacy concerns, and power inequality and how they compare with a number of cultural dimensions (Pace of Life, Individualism, Uncertainty Avoidance and Power Distance).
We find that behavioral  differences exist  across cultures in \ya.
Table~\ref{all-hypotheses} shows a summary of all the hypotheses involving cultural indices and the results found.

\begin{table*}[ht]
\centering
\scalebox{.87}
{
\begin{tabular}{| p{17cm} |p{2.2cm}|}
%\textbf{Name} & Value \\
%\hline
%\hline
\hline
\textbf{Pace of Life} 	&	\textbf{Correlation}\\
%\hline
\emph{Users from countries with a higher Pace of Life score show more temporally predictable activities (asking, answering and reporting)}& $r_q=0.67$*** $r_a= 0.37$**  $r_r=   0.18$\\
\hline
\hline
\textbf{Individualism} 	&	\textbf{Correlation}\\
%\hline
\emph{Users from higher individualism index countries provide more answers} 	& $r=0.46$***\\
%\hline
\emph{Users from countries with higher individualism index contribute more to the community than what they take away from the community} &$r=0.37$**\\
%\hline
\emph{Users from more collective (less individualistic) cultures have higher probability to violate CQA norms} & $r=-0.48$**\\
%\hline
\emph{Users from higher individualism index countries exhibit higher level of concern about their privacy} & NA \\
\hline
\hline
\textbf{Power distance} 	&	\textbf{Correlation}\\

\emph{Users from higher power distance countries show larger indegree imbalance in follow relationships}& $r=0.65$***\\
\hline
\hline
\textbf{Uncertainty Avoidance} 	&	\textbf{Correlation}\\

\emph{Users from countries with higher uncertainty avoidance index exhibit more temporally predictable activities (asking, answering and reporting)} & $r_q=-0.43$** 	$r_a=-0.55$** $r_r=-0.51$**\\
\hline
\end{tabular}
}
\caption{Pearson correlation coefficients in hypotheses related to pace of life, individualism, uncertainty avoidance and power distance.
$p$-values are indicated as: $p$<0.005(***), $p$<0.05 (**), $p$<0.1 (*).}

\label{all-hypotheses}
\end{table*}

We acknowledge that our study is observational and lacks controlled experimental ground truth data.
Therefore, we cannot draw causal conclusions whether cultures shape the ecosystem of \ya.
%Also, there exist high variability across individuals in a country, hence based on the results of this study, we should not stereotype users in Yahoo Answers.
However, our results hint at the importance of culture-aware CQA moderation.
Note that CQA platforms like \ya\ employ human moderators to evaluate reported abuses and determine the appropriate responses, from removing content to suspending user accounts.
We find that collective cultures are more probable to provide bad answers.
At a minimum, more attention of moderators are expected in these cultures to keep the environment clean.

We find that individualistic cultures are more engaged in YA, e.g., by providing more answers and contributing more than their take away.
These results confirm the generalization that individualistic cultures are highly attracted to the Internet.
Researchers often attribute the egalitarian, democratic nature of the Internet to this engagement~\cite{de2000future}.

The evidence of different engagement patterns and difference in pace of life across cultures in CQA platforms imply that some core functionalities such as \emph{question recommendation} and \emph{follow recommendation} could benefit from exploiting cultural factors.
In \emph{question recommendation}, questions are routed to the most appropriate answerers.
To find out such answerers, factors such as followers, interests, question diversity and freshness~\cite{Szpektor2013Reco} are considered.
Our study suggests that including cultural variables such as individualism can be useful.
For example, as users from collective cultures are less probable to answer, questions from those communities should be routed to a larger number of potential answerers.

Another variable, Pace of Life, could also be a factor in \emph{question recommendation}.
Our results show that users from countries with a higher pace of life are temporally more predictable.
In those cultures, if questions are forwarded to answerers during the busy hours of the day (e.g., during office hours), the questions are less likely to get an answer.
Solutions could include routing questions to a larger number of potential answerers, diversifying the set of answerers to include users from countries with a lower Pace of Life, or delaying routing for after work hours.

In  the \emph{follow recommendation}, CQA platforms recommend which other users one can follow based on shared interests, common contacts, and other related factors. 
We find that  in YA, lower power distance countries  show less indegree imbalance in follow relationships.
For follow recommendation in those countries, users to be followed may be recommended to a user with the same level of indegree as them.

CQA platforms could also exploit cultural differentiations to improve targeted ads.
Okazaki and Alonso~\cite{okazaki2003right} analyzed online advertising appeals such as ``soft sell'' appeal (that works by creating emotions and atmosphere via visuals and symbols)  and ``hard sell'' appeal (that provides focus product features, explicit information, and competitive persuasion) across a number of cultures.
They found individualistic cultures like the USA are more attracted to ``hard sell'' appeal, where collective cultures like Japan are attracted to ``soft sell'' appeal.
Ju-Pak's study~\cite{ju1999content} also confirms that fact-based appeal is dominant in the USA, but text-limited, visual layouts are popular in collective cultures like South Korea.
Linguistic aspects in the ads might also be important.
For example, focusing on `I', `me'   in individualistic cultures and `us' and `we' in collective cultures.
Finally, CQA sites could leverage cultural variations in their platforms by, for example, placing textual, informative feature ads to users from individualistic cultures and visual and symbolic ads to users from collective cultures.

\section{Acknowledgments}
The work was funded by the National Science Foundation under the  grant CNS 0952420, and by the Yahoo's Faculty Research and Engagement Program.

\bibliographystyle{abbrv}
\bibliography{Bibtex}  % 

\end{document}